# $In_xGa_{1-x}Sb$ MOSFET: Performance Analysis by Self Consistent CV Characterization and Direct Tunneling Gate Leakage Current

Md. Hasibul Alam[1], Iftikhar Ahmad Niaz[2], Imtiaz Ahmed[2], Zubair Al Azim[2],
Nadim Chowdhury[1] and Quazi Deen Mohd. Khosru[2]
[1,2]Department of Electrical and Electronic Engineering
[1]Bangladesh University of Engineering and Technology, Dhaka-1000, Bangladesh
[2]Green University of Bangladesh, Dhaka-1207, Bangladesh
[1]E-mail: hasib.alam@gmail.com

***Abstract*-In this paper, Capacitance-Voltage (C-V) characteristics and direct tunneling (DT) gate leakage current of antimonide based surface channel MOSFET were investigated. Self-consistent method was applied by solving coupled Schrödinger-Poisson equation taking wave function penetration and strain effects into account. Experimental I-V and gate leakage characteristic for p-channel $In_xGa_{1-x}Sb$ MOSFETs are available in recent literature. However, a self-consistent simulation of C-V characterization and direct tunneling gate leakage current is yet to be done for both n-channel and p-channel $In_xGa_{1-x}Sb$ surface channel MOSFETs. We studied the variation of C-V characteristics and gate leakage current with some important process parameters like oxide thickness, channel composition, channel thickness and temperature for n-channel MOSFET in this work. Device performance should improve as compressive strain increases in channel. Our simulation results validate this phenomenon as ballistic current increases and gate leakage current decreases with the increase in compressive strain. We also compared the device performance by replacing $In_xGa_{1-x}Sb$ with $In_xGa_{1-x}As$ in channel of the structure. Simulation results show that performance is much better with this replacement.**

## I. INTRODUCTION

III-V materials have emerged as a potential candidate since silicon based MOSFETs have reached their fundamental limit. For n-channel device, III-V materials have shown high drive current due to their high electron mobility [1]-[2]. But finding a suitable III-V material with higher hole mobility was a challenge to the research community. Recently hole mobility as high as 1500 $cm^2/Vs$ in strained $In_xGa_{1-x}Sb$ channel has been reported [1]. Antimony based materials also have higher electron mobility as InSb has the highest electron mobility known so far. Moreover, higher cut-off frequencies are also reported for n-channel [3] and p-channel [4] $In_xGa_{1-x}Sb$ FET (Field Effect Transistor) devices. Hence, $In_xGa_{1-x}Sb$ has become a promising candidate for CMOS technology in III-V materials. Although p-channel and n-channel MOSFETs are both under extensive experimental research, a thorough self-consistent study on C-V characteristics and gate leakage current is yet to be done in literature. In this work, we conduct a complete study of C-V characteristics and direct tunneling (DT) gate leakage current using self consistent Schrödinger-Poisson method in $In_xGa_{1-x}Sb$ n-channel MOSFET by varying different process parameters as well as physical parameters like oxide thickness, channel composition, channel thickness and temperature.

## II. SELF CONSISTENT MODELING

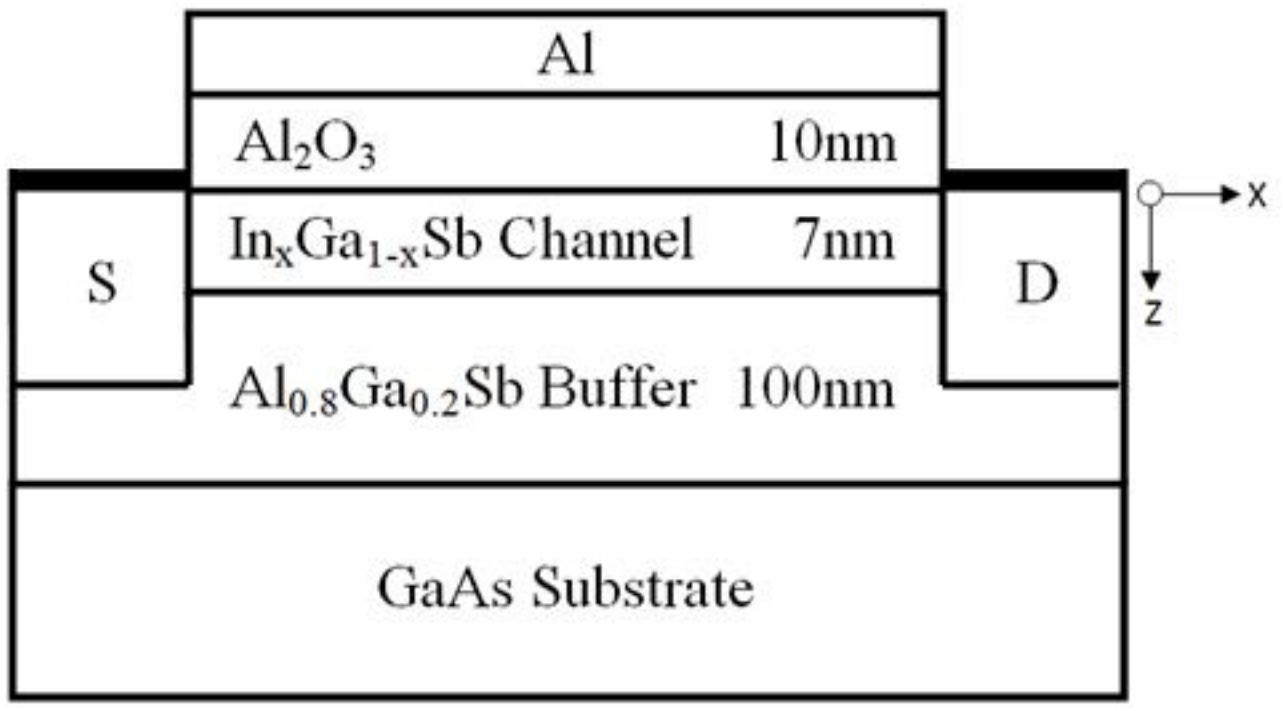

Fig. 1. Basic device structure of $In_xGa_{1-x}Sb$ surface channel MOSFET.

We developed a device simulator to characterize the device in Fig. 1 [5] which is based on self-consistent simulation of Schrödinger-Poisson equation [6]-[7]. In this work, Hamiltonian Matrix formalism [8] is used to solve the Schrödinger's equation numerically. Strain effect is also incorporated in this simulator [9]. Direct Tunneling Gate Leakage Current was obtained using the method described in ref. [10] where transmission probability was calculated using modified WKB (Wentzel–Kramers–Brillouin) method. In our work, we calculated transmission probability using transmission line analogy [11]. Doping concentration of $N_A=10^{17}cm^{-3}$ was used in both channel and buffer region.

The conduction band profile and carrier profile for the device in Fig. 1 are illustrated in Fig. 2.

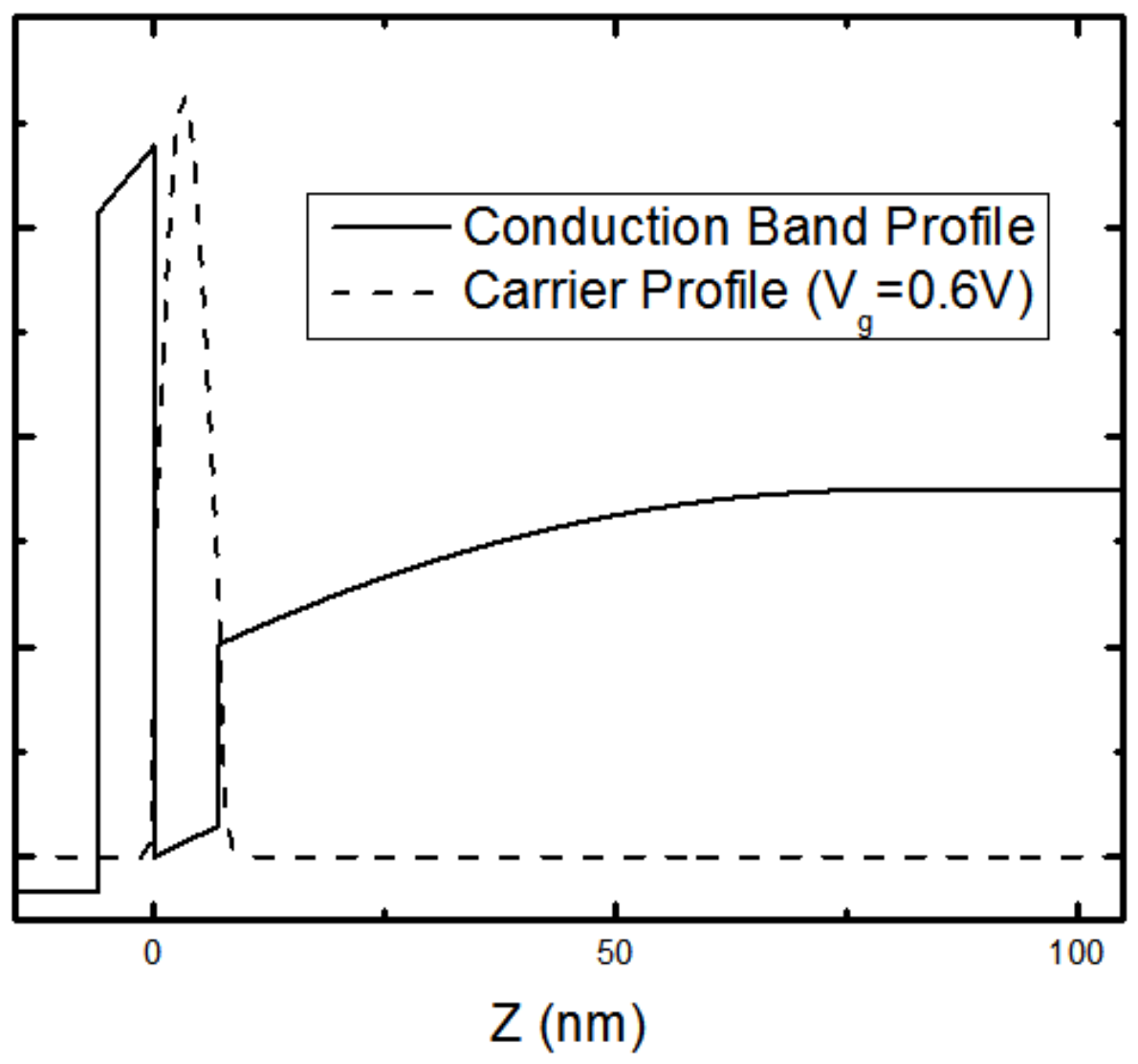

Fig. 2. Conduction band profile along with carrier profile (not drawn to scale).

## III. RESULTS AND DISCUSSIONS

### *A. Effect of Oxide Thickness*

We observed the effect of oxide thickness variation on gate capacitance and DT gate leakage current in Fig. 3 and Fig. 4 respectively.

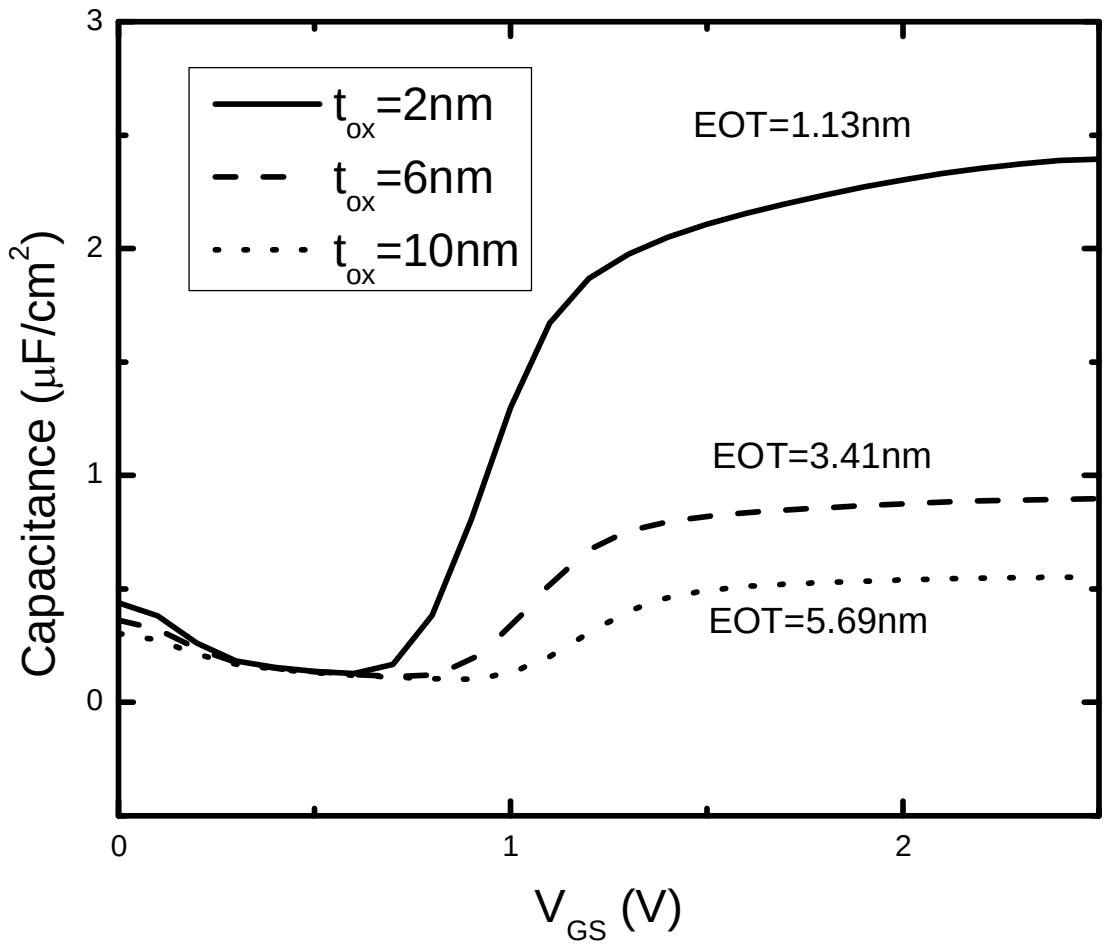

Fig. 3. Gate capacitance as a function of gate voltage for different oxide thickness for $In_{0.35}Ga_{0.65}Sb$ channel at T=300K.

The capacitance is higher for lower oxide thickness. In contrast the threshold value of gate voltage is higher for higher oxide thickness. This can be explained by the sheet carrier density in the oxide semiconductor interface which starts to increase with a lower slope at higher gate voltages for higher oxide thickness.

With higher oxide thickness tunneling probability reduces significantly and hence gate leakage current reduces tremendously.

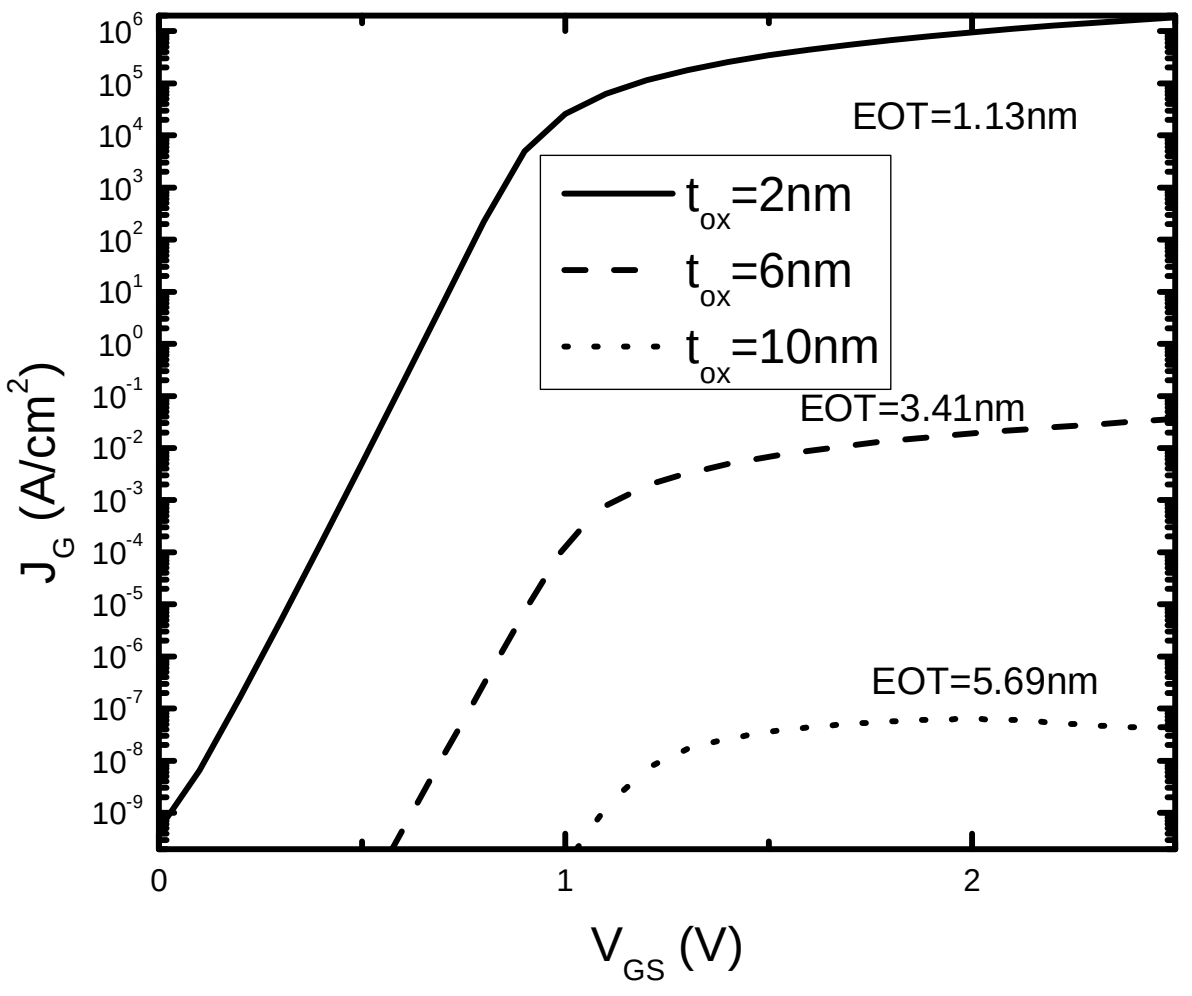

Fig. 4. Direct tunneling gate leakage current as a function of gate voltage for different oxide thickness for $In_{0.35}Ga_{0.65}Sb$ channel at T=300K.

### *B. Strain Effects*

Effect of strain between channel and buffer region on gate capacitance is illustrated in Fig. 5.

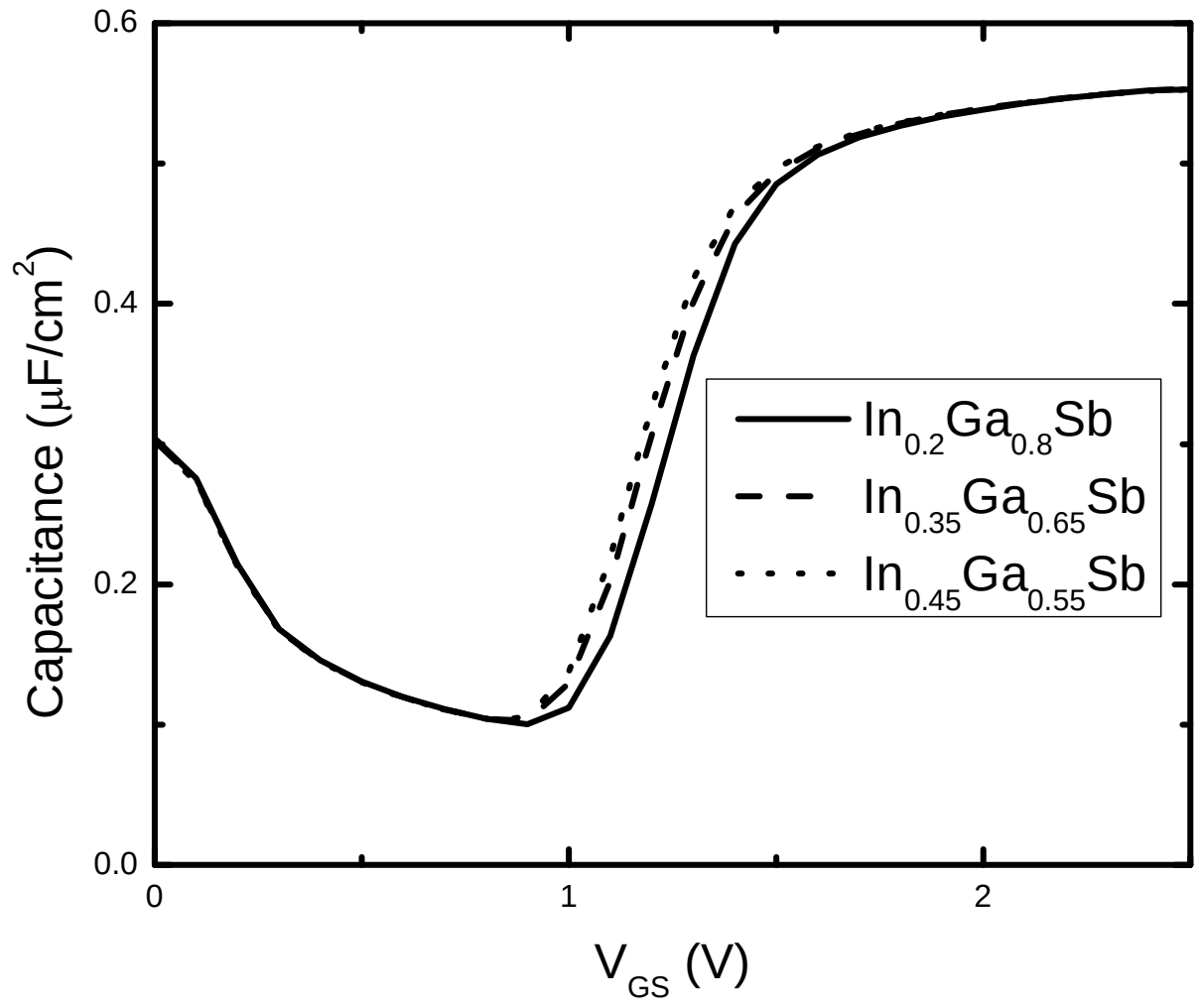

Fig. 5. Gate capacitance as a function of gate voltage for different compositions of GaSb and InSb at channel.

The inversion carrier density increases at a higher rate for higher compressive strain. Hence gate capacitance is higher for higher compressive strain in the moderate inversion region. However, at strong inversion the slope of sheet carrier density becomes same for all three compositions of InSb and GaSb. Hence the gate capacitance becomes the same at strong inversion.

The DT Gate leakage current is shown in Fig. 6. We found the highest value in the lowest compressively strained channel in strong inversion. This can be explained by the higher transmission probabilities (TP) for lower eigen energies in lower compressively strained channel. The contrast behavior before the cross-over can also be justified by the same reasoning where the TPs are lower for lower eigen energies in higher compressively strained channel.

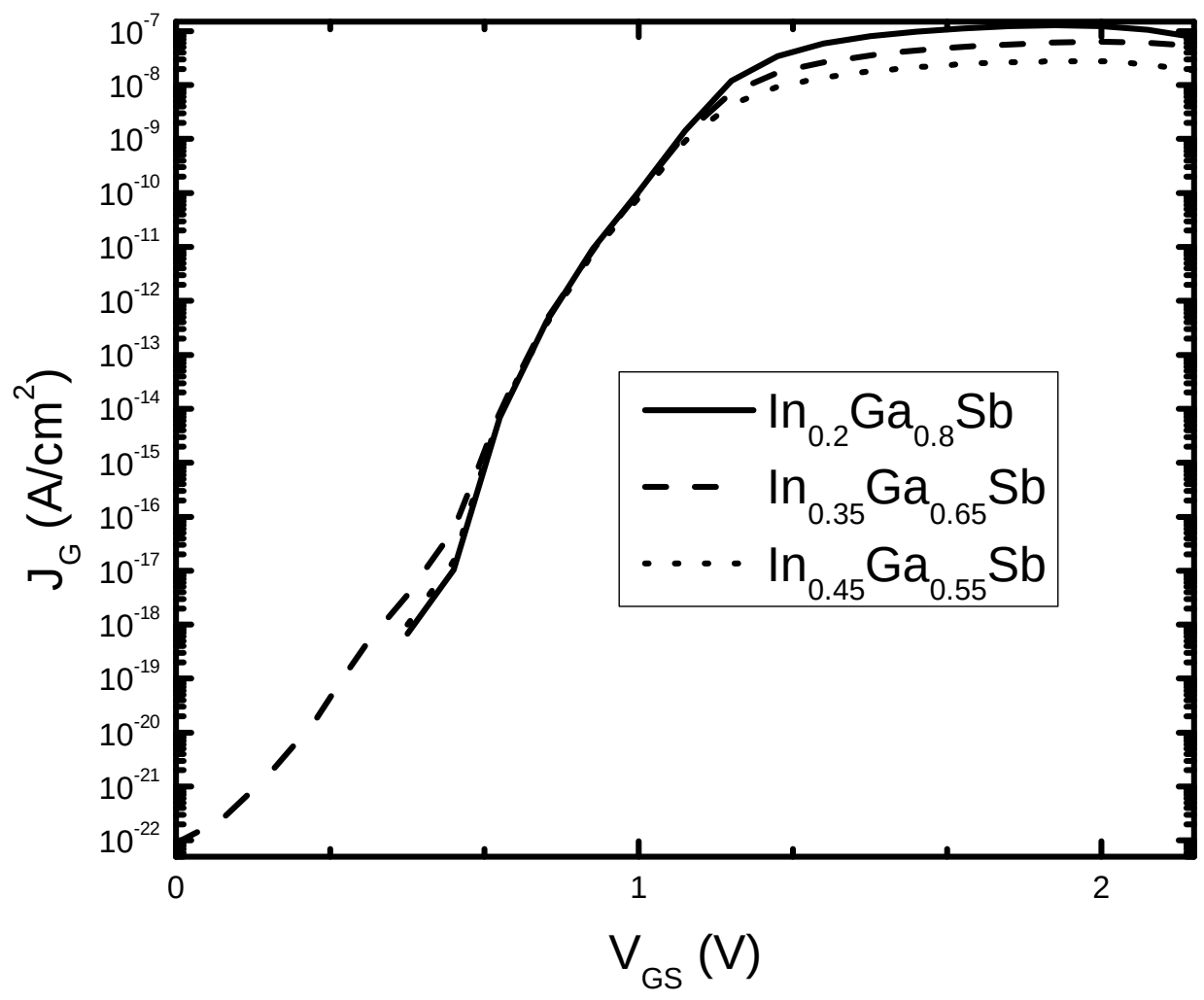

Fig. 6. Direct tunneling gate leakage current as a function of gate voltage for different compositions of GaSb and InSb at channel.

### *C. Effect of Channel Thickness Variation*

Fig. 7 shows the C-V characteristics for different channel thicknesses. There is almost negligible effect of channel thickness on gate capacitance before threshold and at strong inversion. However at moderate inversion the capacitance is slightly higher for higher channel thickness. This is because in wider channel the eigen energies are less distantly spaced compared to narrower channel. As a consequence the carrier confinement is better in wider channel region which gives rise to higher gate capacitance.

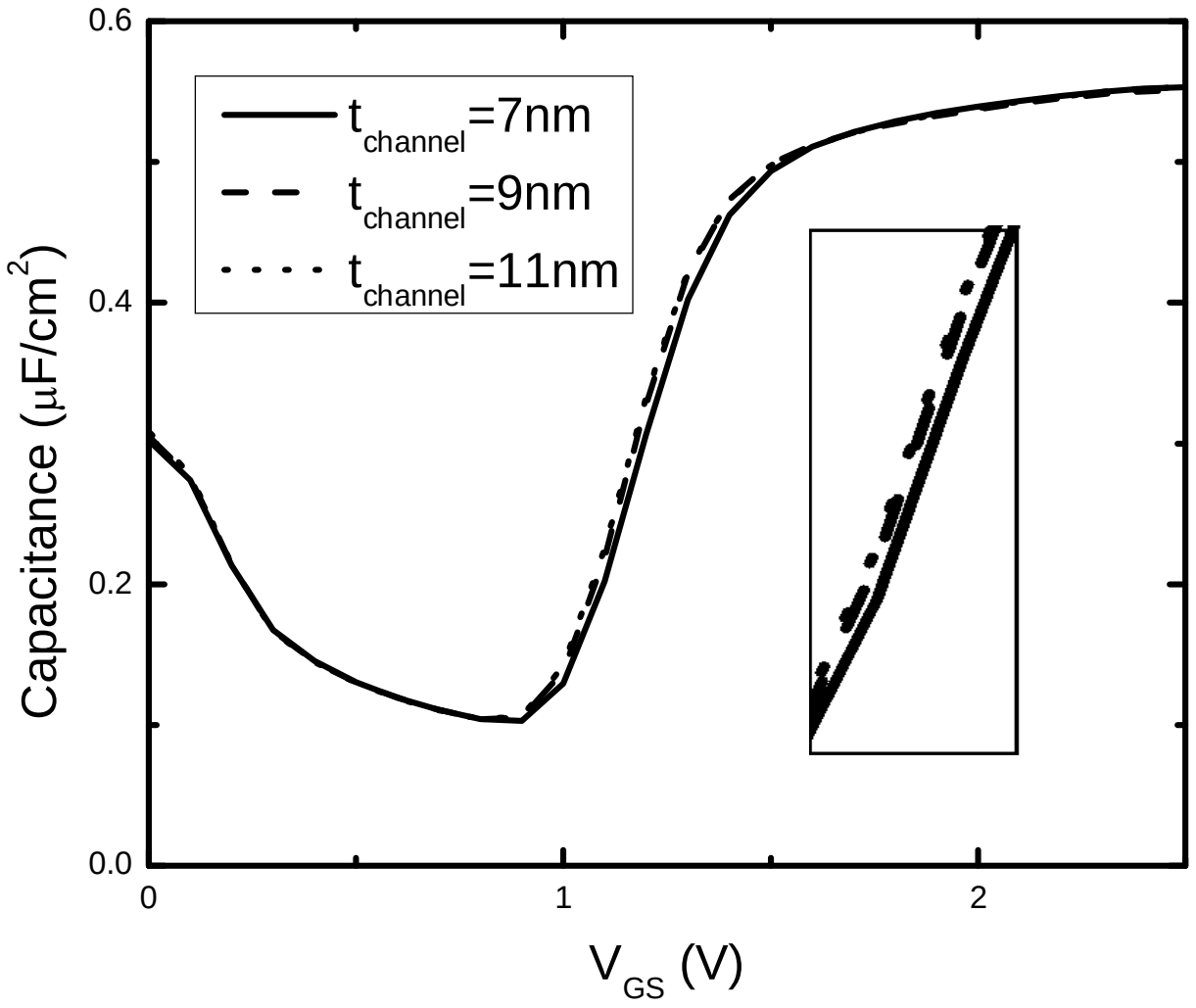

Fig. 7. Gate capacitance as a function of gate voltage for different channel thickness.

The gate leakage behavior is also relatively independent on channel thickness as shown in Fig. 8.

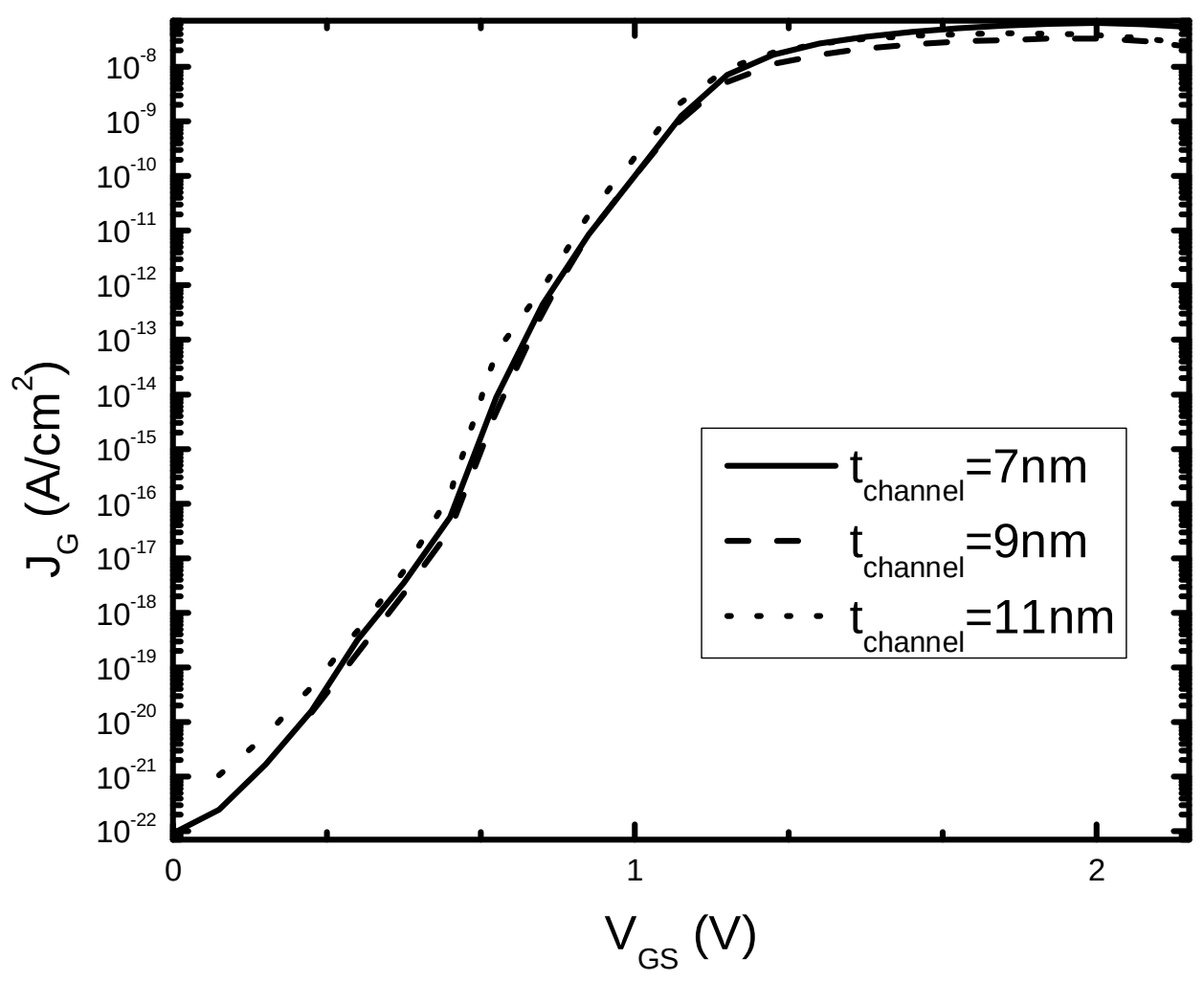

Fig. 8. Direct tunneling gate leakage current as a function of gate voltage for different channel thickness.

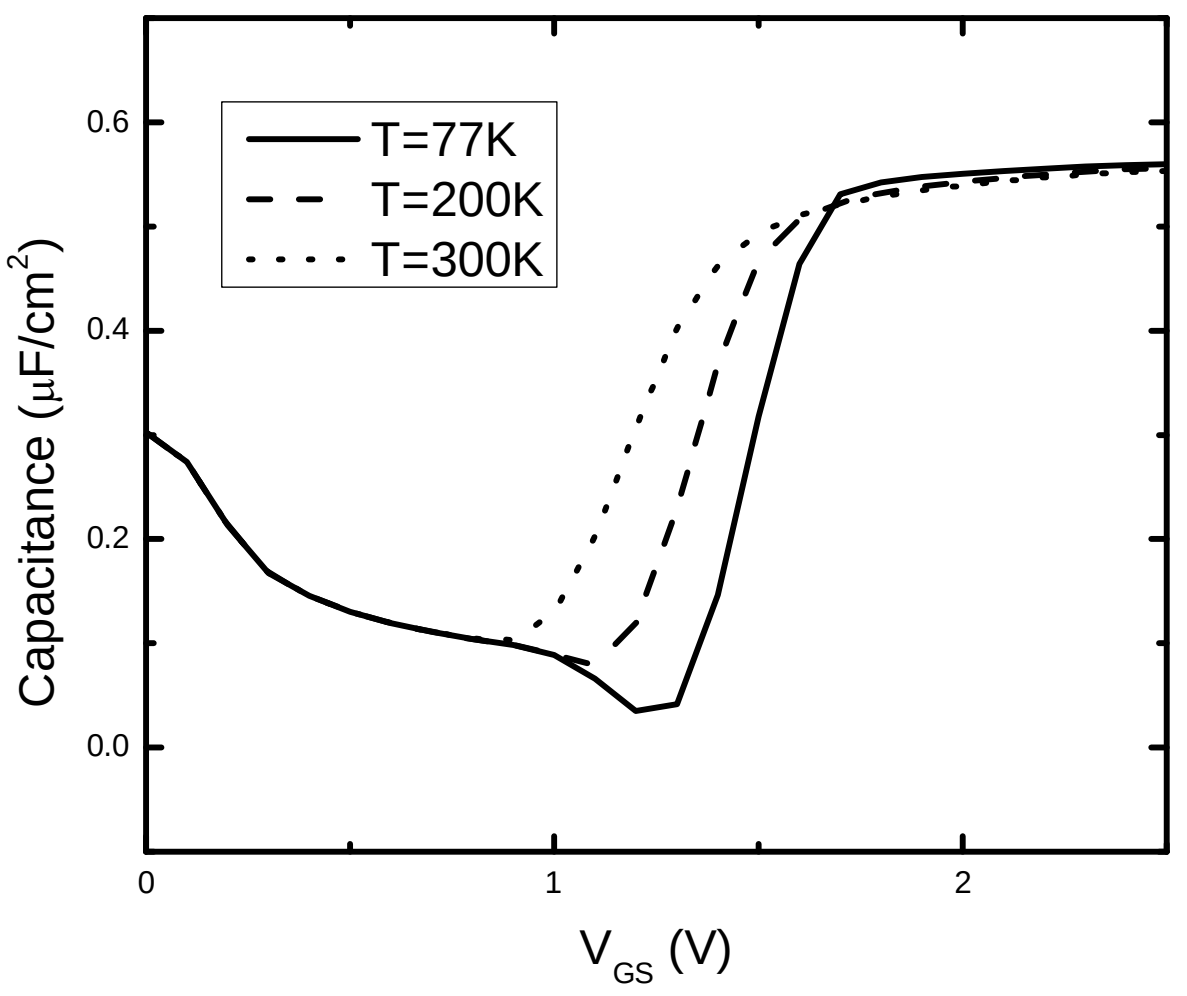

Fig. 9. Gate capacitance as a function of gate voltage for different temperatures.

The slight variation occurs due to the variation of product of TP with carrier density for different channel thicknesses at different voltages.

## *D. Effect of Temperature*

The gate capacitance, $1^{st}$ eigen energy, channel sheet carrier density and DT gate leakage current for three temperatures are illustrated in Fig. 9, Fig. 10, Fig. 11 and Fig. 12 respectively. The gate capacitances before the onset of inversion are same because of the same carrier occupancy in the $1^{st}$ Eigen state. At higher gate voltages $1^{st}$ eigen energy is higher and carrier occupancy is lower for $1^{st}$ subband minima at higher temperature. But additional carrier from higher subbands makes the capacitance higher at higher temperature.

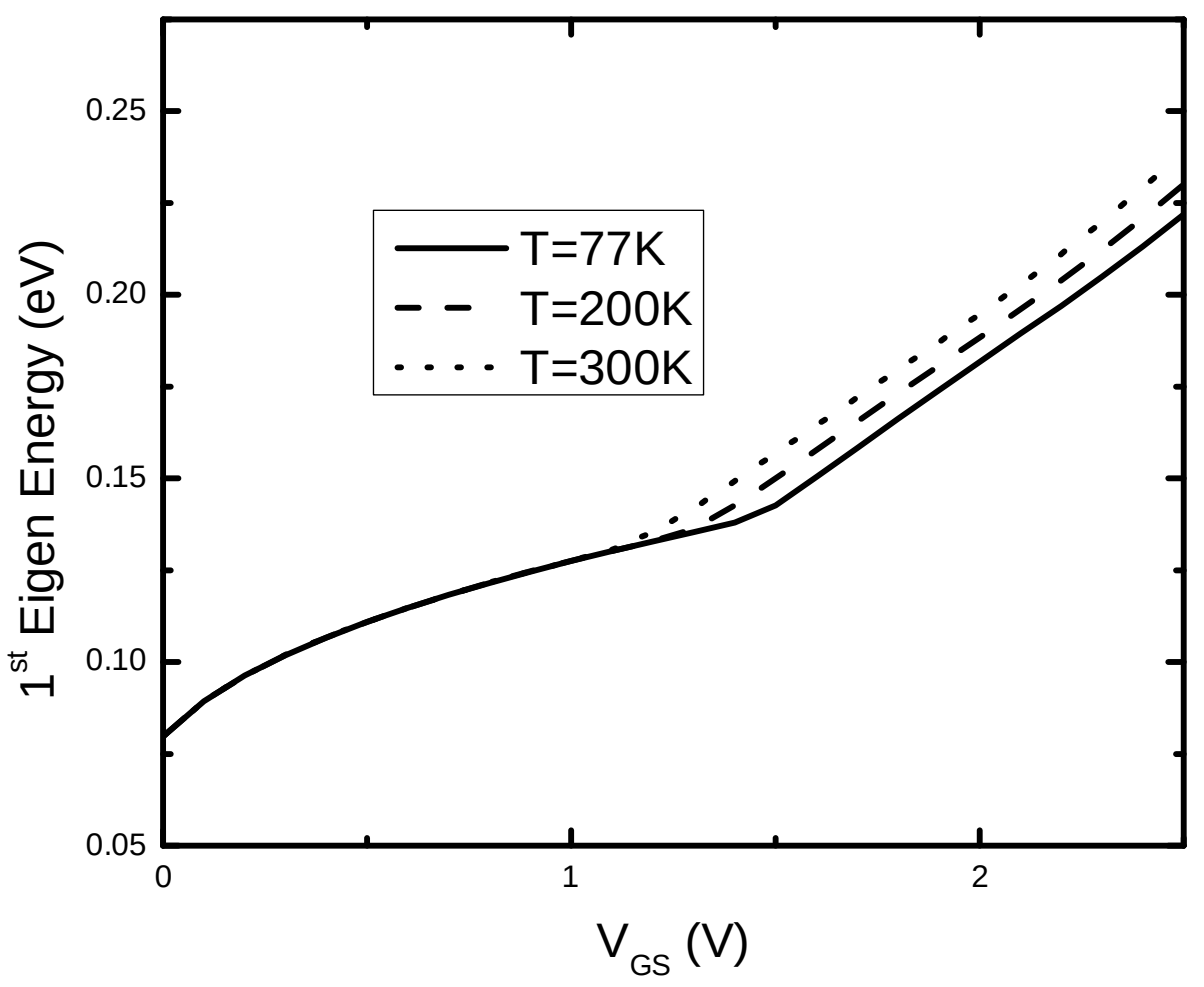

Fig. 10. $1^{st}$ Eigen Energy as a function of gate voltage for different temperatures.

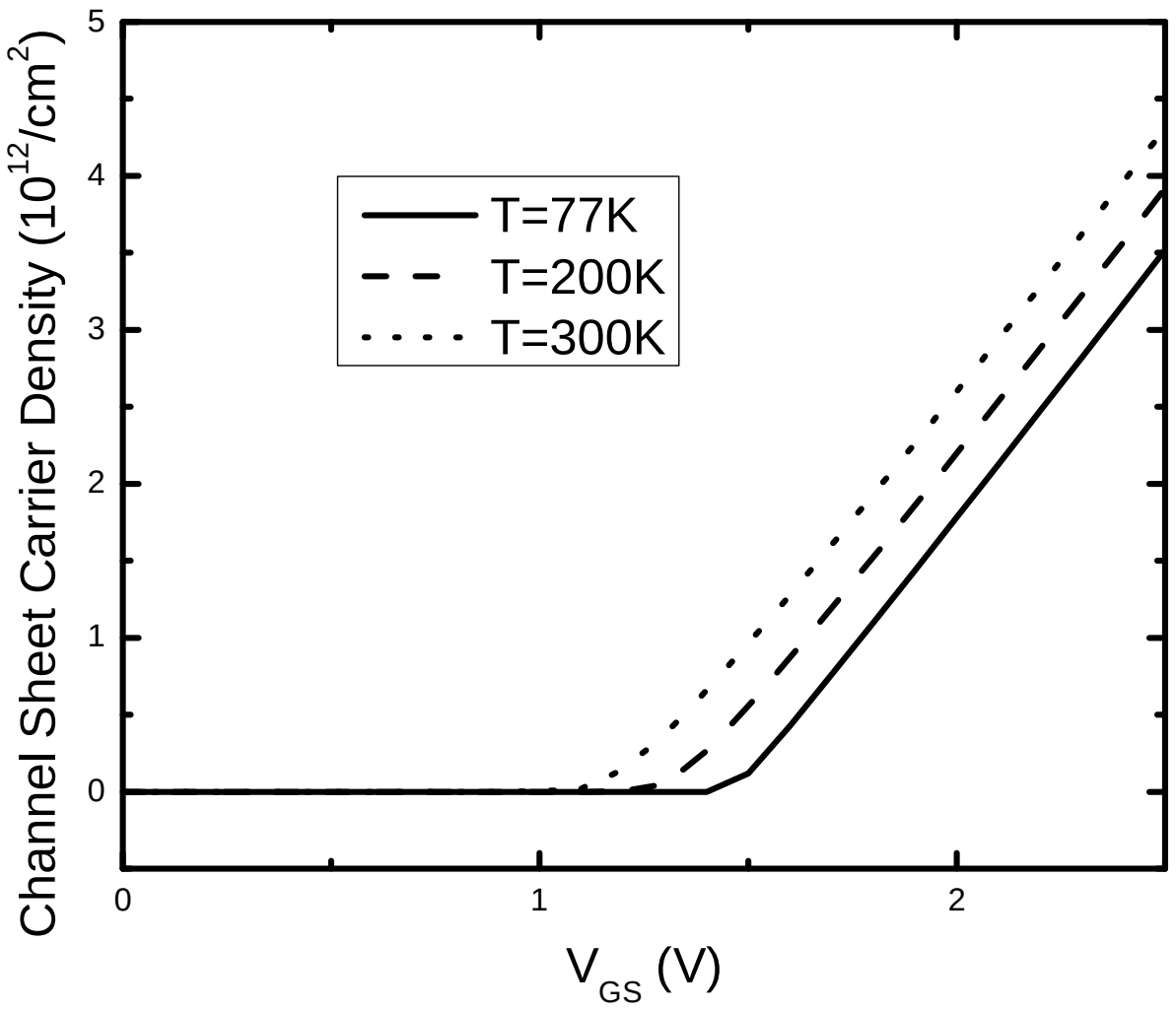

Fig. 11. Channel Sheet Carrier Density as a function of gate voltage for different temperatures.

DT gate leakage current shows an interesting behavior. For T=300K it is highest at all gate voltages. But for T=77K & 200K there is a cross over. This can be attributed to the fact that for T=300K TP is highest for all gate voltages. For T=77K, TP is lower for all eigen energies at voltages before the cross over whereas for T=200K TP exhibits the opposite behavior. The sheet carrier density increases linearly after threshold voltage with the highest value for T=300K and lowest for T=77K. However the product of carrier density and TP becomes higher in case of T=77K which explains higher gate leakage current compared to T=200K.

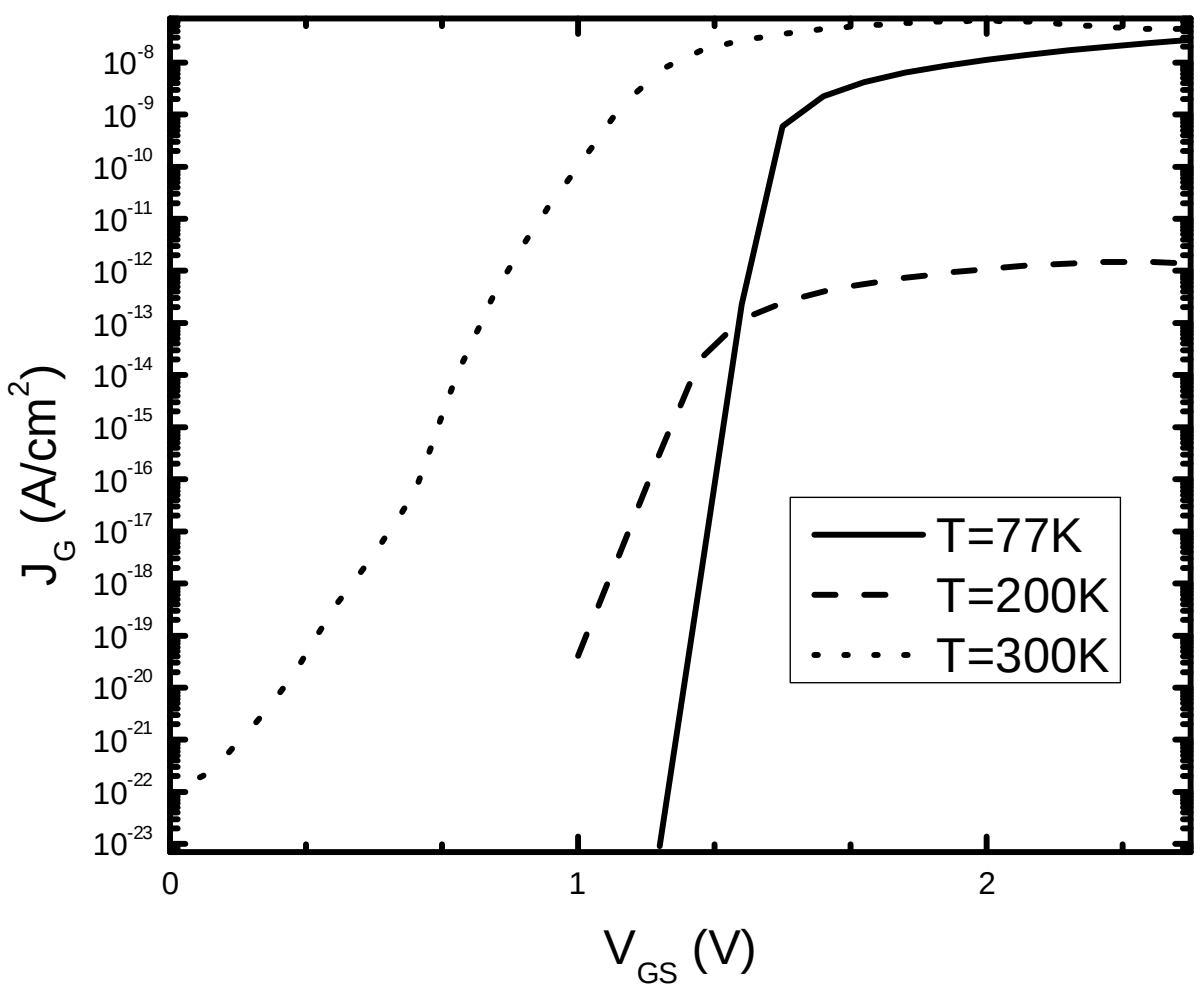

Fig. 12. Direct tunneling gate leakage current as a function of gate voltage for different temperatures.

We also investigated ballistic drain current [12]-[13] to observe the effect of strain which is illustrated in Fig. 13.

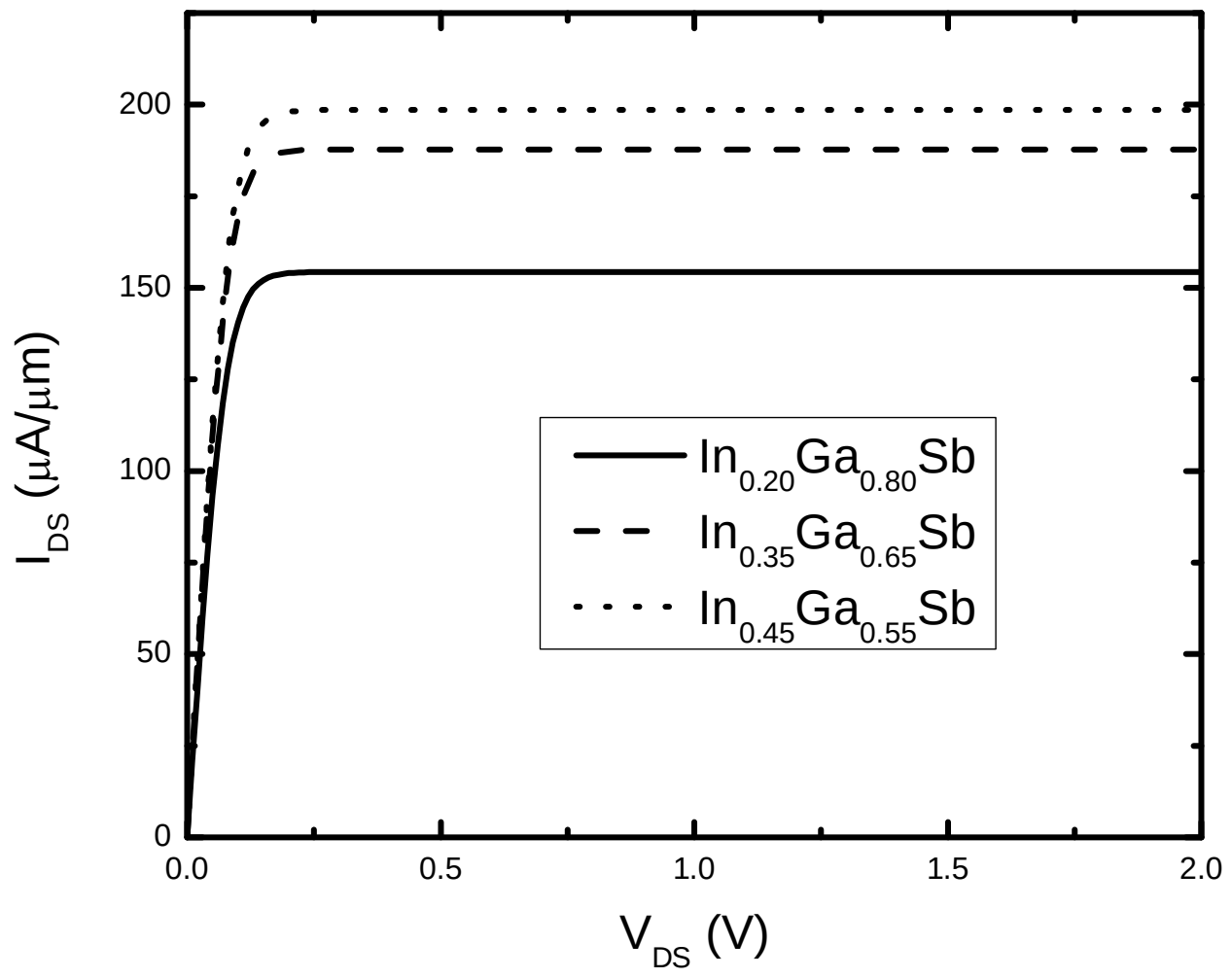

Fig. 13. Drain current as a function of drain voltage for different channel compositions ($V_{GS}$=1.6V).

## *E. Performance Comparison with InGaAs MOSFET*

We compared gate capacitance, gate leakage current and ballistic drain current of InGaSb with those of InGaAs (same device dimensions and doping, only channel and buffer region replaced by $In_{0.75}Ga_{0.25}As$ and $In_{0.53}Ga_{0.47}As$ respectively) which are illustrated in Fig. 14, Fig. 15 and Fig. 16 respectively.

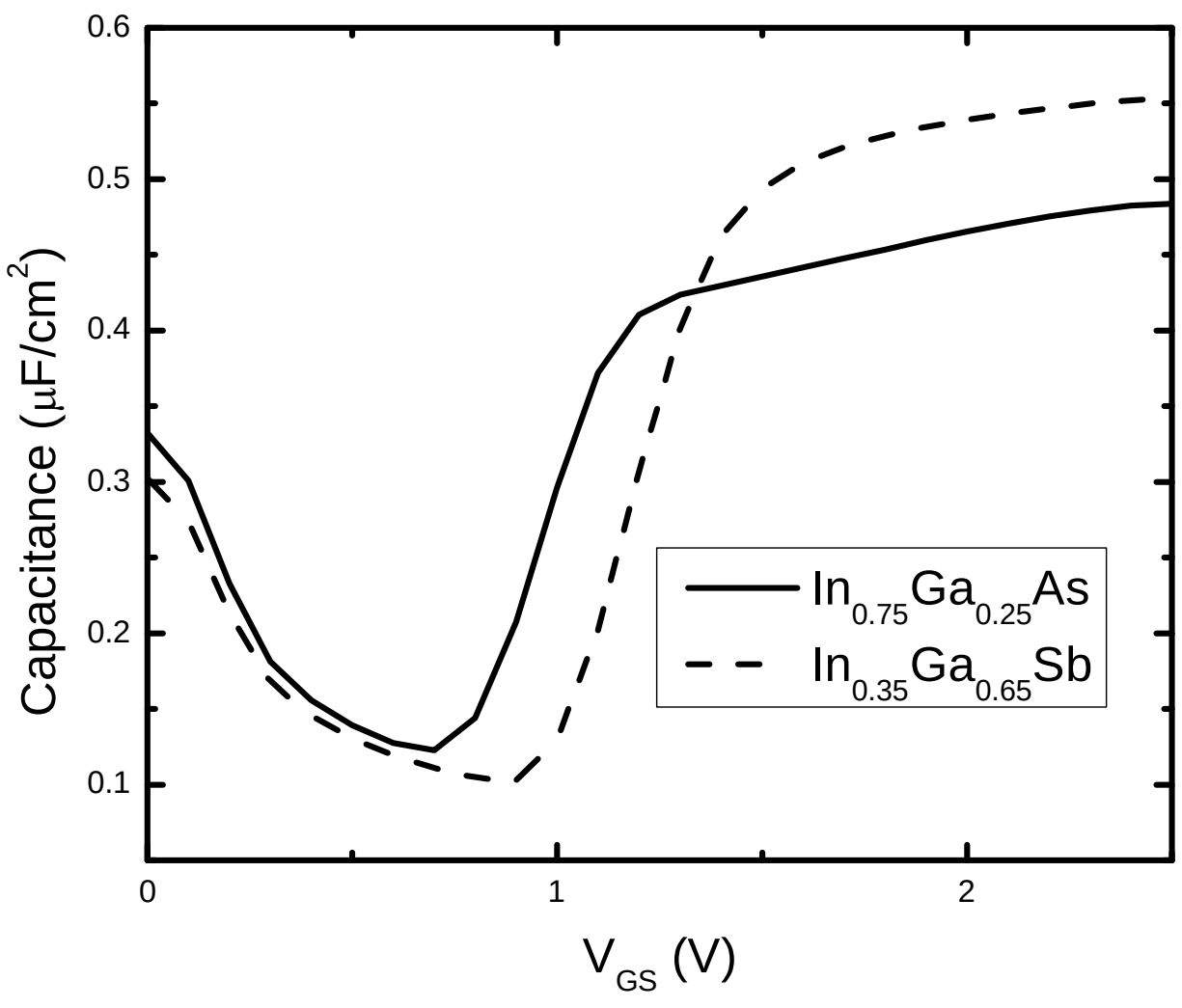

Fig. 14. Gate capacitance as a function of gate voltage for two MOSFET structures.

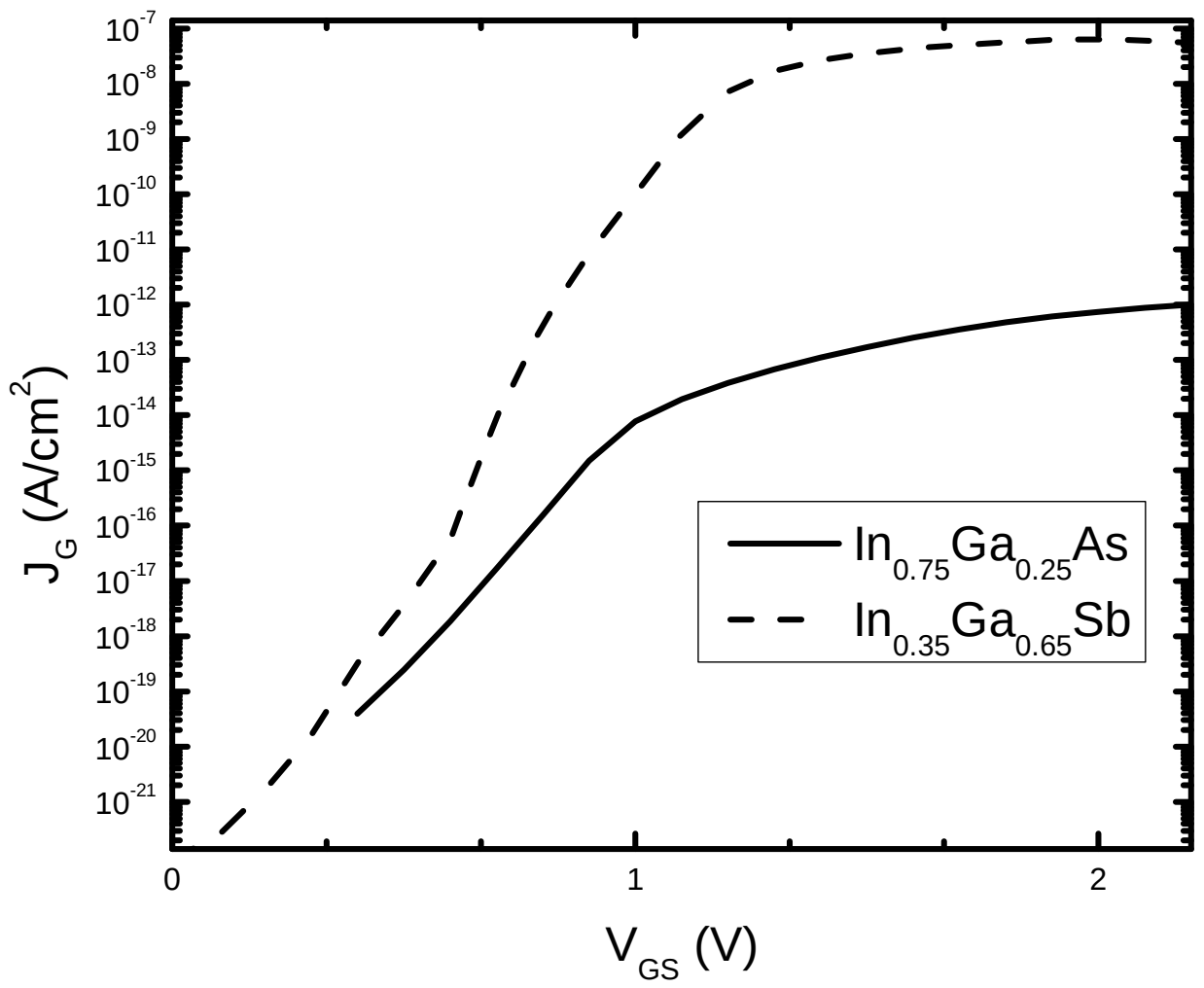

Fig. 15. Direct tunneling gate leakage current as a function of gate voltage for two structures.

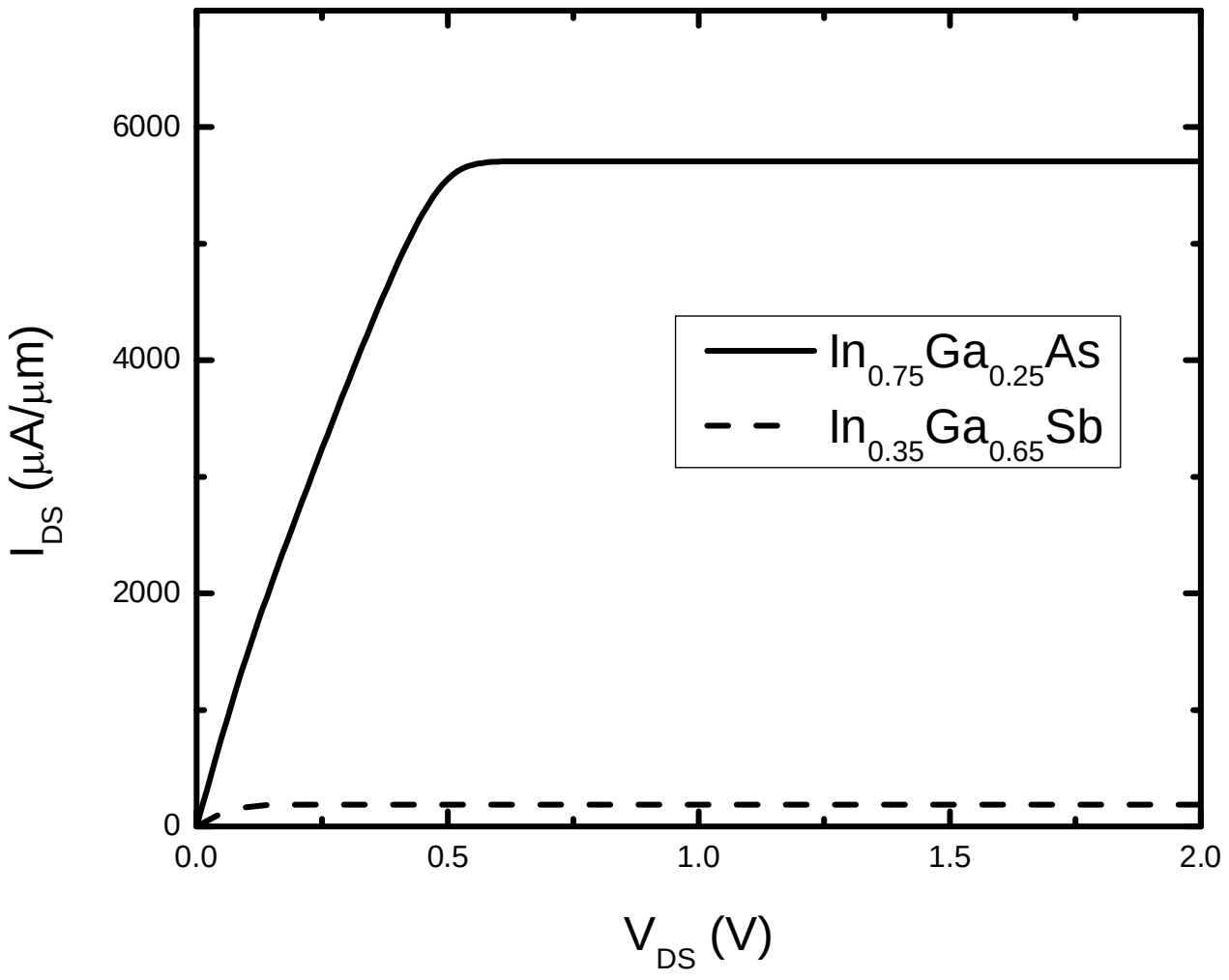

Fig. 16. Drain current as a function of drain voltage for two structures ($V_{GS}$=1.6V).

From the figures it is obvious that the capacitance is less in InGaAs which means the control on channel is less. However in terms of gate leakage and ballistic drain current the performance of InGaAs is far superior to InGaSb. To ensure the fact we performed all the analyses for different compositions, EOT, channel thickness and all yield the same outcome of performance degradation in InGaSb MOSFETs.

## IV. CONCLUSIONS

We investigated C-V characteristics and gate leakage current with self-consistent method for $In_xGa_{1-x}Sb$ surface channel MOSFET. We observed that increase of compressive strain has negligible effect on inversion capacitance but it significantly reduces leakage current and improves ballistic performance. Variations of temperature and oxide thickness also have strong effects on C-V characteristics and gate leakage currents whereas channel thickness variation has little effect. Our simulated results show that n-channel $In_xGa_{1-x}As$ MOSFET is far better than its counterpart $In_xGa_{1-x}Sb$ in terms of gate leakage current and ballistic performance.

## ACKNOWLEDGMENT

The authors are thankful to Mr. Raisul Islam for his help in simulation.